\documentclass[reprint,prb,amsmath,amssymb,aps]{revtex4-1}

\usepackage{graphicx}% Include figure files
\usepackage{dcolumn}% Align table columns on decimal point
\usepackage{bm}% bold math
\usepackage{longtable}
\usepackage{array}

\begin{document}

\title{Ab initio investigation of FeAs/GaAs heterostructures for potential
spintronic and superconducting applications}

\author{Sin\'{e}ad M. Griffin}
\email{sgriffin@ethz.ch}
\homepage{http://sites.google.com/sineadv0}
\affiliation{Materials Department, University of California, Santa Barbara, California 93106-5050, USA}
\affiliation{Department of Materials, ETH Zurich,
Wolfgang-Pauli-Strasse 27, CH-8093 Zurich, Switzerland}

\author{Nicola A. Spaldin}
\email{nicola.spaldin@mat.ethz.ch}
\homepage{http://www.theory.mat.ethz.ch}
\affiliation{Department of Materials, ETH Zurich,
Wolfgang-Pauli-Strasse 27, CH-8093 Zurich, Switzerland} 

\date{\today}

\begin{abstract}
Ultra-thin FeAs is of interest both as the active component in the newly
identified pnictide superconductors, and in spintronic applications at the
interface between ferromagnetic Fe and semiconducting GaAs. Here we use
first-principles density functional theory to investigate the properties
of FeAs/GaAs heterostructures. 
We find that the Fermi surface is modified from that characteristic of
the pnictide superconductors by interactions between the FeAs layer and
the As atoms in the GaAs layers. 
Regardless of the number of FeAs layers, the Fe to As ratio, or the strain
state, the lowest energy magnetic ordering is always antiferromagnetic, suggesting
that such heterostructures are not promising spintronic systems, 
and offering an explanation for the failure of spin injection across Fe/GaAs 
interfaces.

\end{abstract}

%\keywords{????????latex-community, revtex4, aps, papers}

\maketitle

\section{Introduction}

FeAs-based materials have attracted much attention in recent years, first due to 
their potential as magnetic semiconducting systems \cite{Prinz:1990, Munekata:1989, Ohno:1996, Shirai:2001, Rahman:2006}, and more recently for their 
unexpected high-$T_{\rm c}$ superconductivity \cite{Hosono:2008}. 

Magnetic semiconducting systems are of interest because they could
in principle enable so-called ``spintronic'' devices that exploit 
the spin degree of freedom of the electron as well as its charge \cite{Prinz:1998}.
One route to spintronic behavior is through hybrid structures, in which 
magnetic metals are used to inject spin-polarized electrons 
into semiconductors \cite{Ohno:1998}. Here the Fe/GaAs system is particularly appealing 
because of the high ferromagnetic Curie temperature of Fe, the 
well-established semiconducting properties of GaAs, and the close 
lattice match between body-centered cubic Fe and zincblende GaAs. 
However spin injection in Fe/GaAs has not yet proven successful \cite{Krebs:1982}.
One possible reason for this is the formation of other phases at the
interface; indeed FeAs and Fe$_{2}$As have both been reported experimentally \cite{Ruckman:1986, Rahmoune:1997, Lepine:1998, Palmstrom:2002},
and {\it ab initio} calculations suggest that Fe penetrates the GaAs lattice,
breaking the Ga-As bonds in favor of Fe-As and elemental Ga \cite{Mirbt:2003}. 

The recent and entirely unanticipated discovery of superconductivity at 
$\sim$ 50K in layered rare earth oxide / iron arsenide compounds has generated
tremendous excitement 
within the condensed matter community \cite{Wang:2008, Norman:2008}. Previously, all known conventional high temperature 
(high-$T_{\rm c}$) superconductors contained copper-oxygen layers; in fact the
presence of oxygen and the
absence of magnetism were believed to be requirements for high-$T_{\rm c}$ 
behavior \cite{Matthias:1953, Matthias:1955}. In the new materials, however, superconductivity occurs in the magnetic
iron-arsenic planes, in complete violation of previous understanding;
they therefore provide an invaluable new handle for exploring the
long-sought-after mechanism underlying high temperature superconductivity. 
Density functional calculations have revealed correlations between the superconducting
Curie temperature and the normal-state structural and electronic properties of the 
fluorite-structure Fe-As layer. In particular, the out-of-plane Fe-As bond length,
the striped antiferromagnetic order of the undoped parent compound, and the Fermi 
surface nesting that occurs between bands derived from Fe-\textit{d} states seem to be 
important.

In this work we use first-principles density functional theory to calculate the structural
stabilities and electronic properties of a range of FeAs/GaAs superlattices.
The goal of our work is two-fold: First to search for superlattices within this 
family with desirable spintronic properties such as half-metallicity or ferromagnetism,
and second to explore whether the signature electronic properties of the FeAs layers in the
pnictide superconductors can be reproduced in this artificial system.
Our model heterostructures consist of $n$ layers ($n$ = 1, 3) of zincblende GaAs 
alternating with $m$ layers ($m$ = 1, 3) of FeAs in either the zincblende structure
or the antifluorite structure found in the FeAs superconductors. 

\section{Computational Details}

Our density functional theory (DFT) calculations were performed using the Vienna ab initio 
Simulation Package (VASP) \cite{VASP1, VASP2}. 
We expanded the electronic wave functions and density using a plane-wave basis set, and used 
the supplied VASP PAW potentials \cite{PAW} with the Perdew-Burke-Ernzerhog (PBE) exchange-correlation 
functional\cite{PBE1, PBE2} for core-valence separation. A 10$\times$10$\times$10 Monkhorst-Pack \cite{Monkhorst_Pack} \textit{k}-point mesh with a Gaussian smearing of 0.2 eV was used for the Brillouin Zone integrations; these are suitable values for metals. The plane-wave cut-off was set to 400 eV, and  for structural relaxations we allowed the ions to relax until the Hellmann-Feynman forces were less than 1 meV/\AA$^{-1}$̊.

We treated the exchange-correlation functional within the spin-polarized generalized gradient approximation plus Hubbard $U$ 
(GGA+$U$) method \cite{Dudarev} to account for electron correlations and chose a $U$ of 0.5 eV as implemented in the Dudarev scheme.
While there has been much discussion in the literature of the relative appropriateness of 
various functionals (hybrids, LDA$+U$, GGA$+U$, etc.) and $U$ parameters, we find that
GGA+$U$ with $U$=0.5 eV gives good agreement with both hybrid functional calculations and
experimental structural properties for bulk \textit{MnP}-type FeAs\cite{me1:2011}. In
common with all functionals that have been tested to date, the local magnetic moments
are over-estimated compared with those reported experimentally \cite{Mazin:2008}; it remains a matter
of debate whether this is a consequence of the neglect of spin fluctuations in the
density functional formalism \cite{Yin:2008} and/or the difficulties associated with rigorously 
defining a local moment experimentally in such a broad-band metallic system \cite{Walters:2009}.

\section{Results -- Bulk Properties of F\lowercase{e}A\lowercase{s} and F\lowercase{e}$_{2}$A\lowercase{s}}

\subsection{Properties of bulk FeAs in MnP and ZnS structural variants}

We begin with a comparison of bulk FeAs in its experimentally observed ground state 
\textit{MnP} structure\cite{Selte:1969}, and the zincblende structure that is of interest for our spintronic 
superlattice calculations.

\begin{figure}
 \centering
\includegraphics[scale=0.5,bb=0 0 1973 1081,width=8cm]{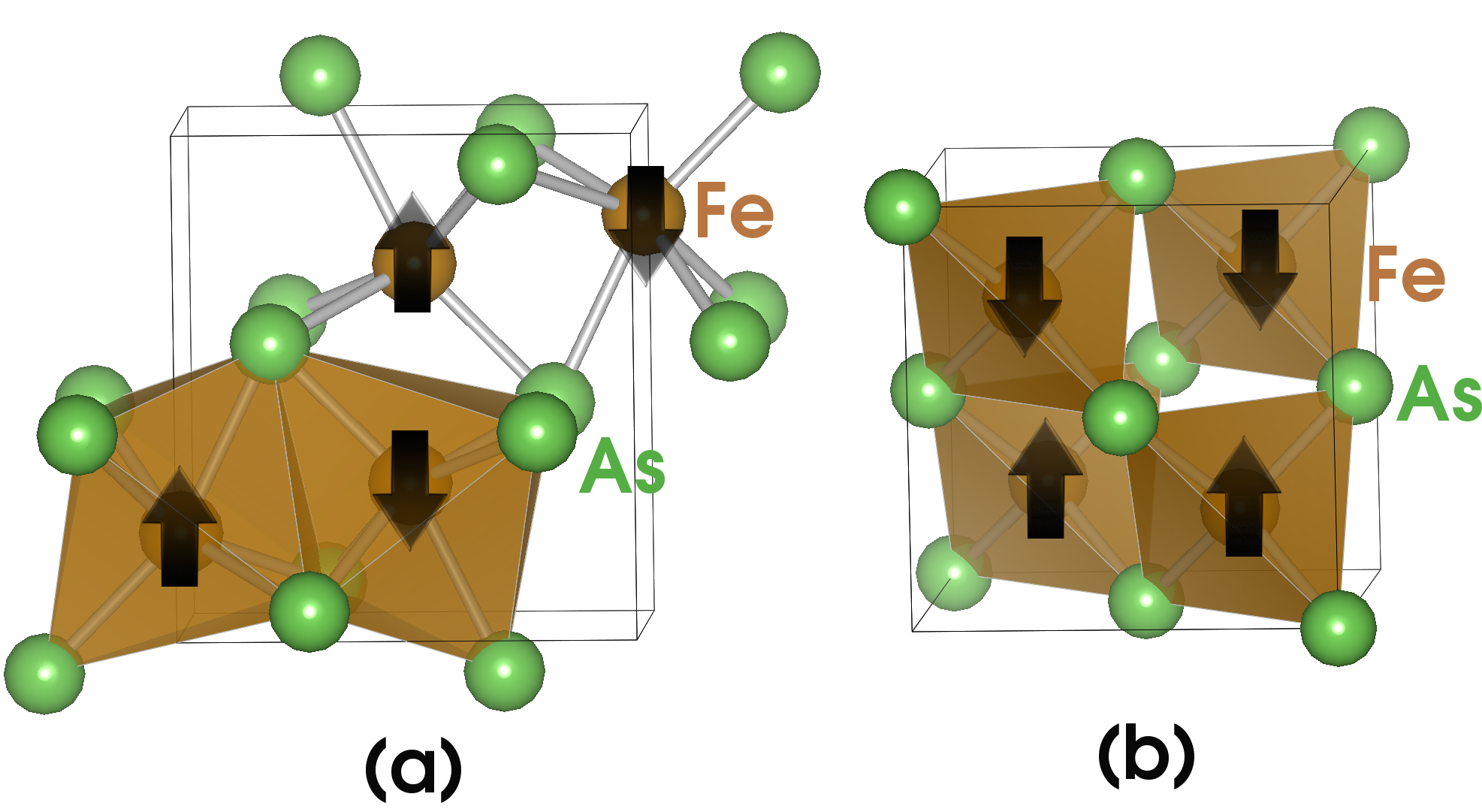}
 % FeAs_vesta.png: 0x0 pixel, 0dpi, 0.00x0.00 cm, bb=
 \caption{(a) Structure of \textit{MnP}-type FeAs with ground state AFM ordering indicated. (b) Structure of zincblende FeAs with arrows indicating ground state AFM ordering.}
 \label{Fig. 1}
\end{figure}

Fig. 1 shows the structures of the \textit{MnP}-type and zincblende FeAs. The former (1(a))
consists of octahedrally coordinated Fe ions with the octahedra edge-shared. The octahedra
are distorted with the Fe ions shifted from their centres. In both cases the 
magnetic ground state is antiferromagnetic indicated by the arrows, with each Fe being antiferromagnetically coupled to all its Fe nearest neighbours.

Zincblende FeAs (1(b)) consists of interpenetrating face-centered cubic sublattices of Fe and As  
that are shifted by ($\frac{1}{4}, \frac{1}{4}, \frac{1}{4}$) along the (111) direction relative to each other.
As a result, both the Fe and As atoms are tetrahedrally coordinated, with corner-sharing polyhedra, and the packing density is lower than in the \textit{MnP} case.

The calculated lattice parameters for \textit{MnP} FeAs are given in Table I.
Our calculations correctly obtain the AFM-ordered \textit{MnP} structure as the ground state,
with the zincblende structure 1.39 eV per 2-atom formula unit higher in
energy. Fig. 2(a) shows the relative stabilities of \textit{MnP}-type and zincblende FeAs as a function 
of volume of a 2-atom formula unit, set by uniformly scaling the equilibrium lattice parameters. The \textit{MnP}-type structure is the ground state until an expanded volume of ~$40$ \AA$^{3}$ per 2-atom
formula unit. At the equilibrium lattice volume of GaAs ($45 $ \AA$^{3}$), the zincblende structure 
is the lowest energy structure, suggesting that it could be the stable phase in coherently grown 
GaAs/FeAs heterostructures, although our calculated lattice constant for zincblende FeAs --
$5.36$\AA\ -- is smaller than that of GaAs. 

\begin{figure}
 \centering
 \includegraphics[scale=0.5,bb=0 0 710 607, width=8cm, keepaspectratio = true]{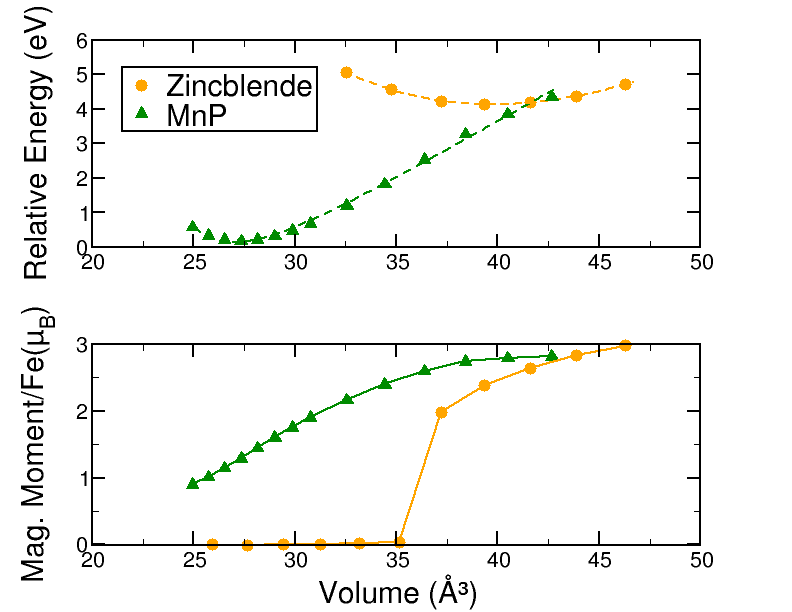}
 % tmnpzns.png: 0x0 pixel, 0dpi, 0.00x0.00 cm, bb=
 \caption{(a) Calculated energy-volume curves for \textit{MnP}-type and zincblende FeAs with AFM ordering. The volume shown is for one 2-atom formula unit of FeAs. The \textit{MnP} structure is stable over a large range of volumes. (b) Absolute value of magnetic moment per Fe for the \textit{MnP} and zincblende structures with AFM ordering as a function of volume.}
 \label{Fig. 2}
\end{figure}

To explore the spintronic properties, we calculated the spin-polarized density of states for 
hypothetical ferromagnetically ordered zincblende FeAs at our calculated equilibrium volume 
(38.5 \AA$^3$),
and at the experimental volume of GaAs ($45$ \AA$^{3}$) (Fig. 3). As expected, we find a narrowing
of the bands as the volume is increased. 
In addition, the exchange-splitting between majority and minority bands is more pronounced at
the expanded volume, resulting in almost half-metallic behaviour, similar to that previously 
reported for zincblende MnAs.\cite{Sanvito:2000}. It is likely that half-metallicity will be 
achieved with the additional band narrowing provided by quantum confinement in thin layers
of FeAs in heterostructures.

\begin{center}
% use packages: array
\begin{table*}
\caption{\label{1} Calculated and experimental crystal structure parameters for \textit{MnP}-type FeAs and \textit{Cu$_{2}$Sb}-type Fe$_{2}$As.} 
\begin{tabular}{|ll||c|c|c||c|c|c|c|}
\hline
 & & a(\AA) & b(\AA) & c(\AA) & \textit{Fe(x)} & \textit{Fe(z)} & \textit{As(x)} & \textit{As(z)} \\ \hline
\textbf{FeAs} & GGA+U & 5.471 & 3.276 & 6.050 & 0.002 & 0.202 & 0.201 & 0.573 \\ 
& Exp\cite{Selte:1969} & 5.442 & 3.372 & 6.028 & 0.0027 & 0.1994 & 0.1993 & 0.5774 \\ 
\textbf{Fe$_{2}$As} & GGA+U & 3.627 & 3.627 & 5.980 & - & 0.329 & - & 0.266 \\ 
& Exp\cite{Adachi:1988} & 3.627 & 3.627 & 5.981 & - & 0.318 & - & 0.266 \\ \hline
\end{tabular}
\end{table*}
\end{center}

Finally, to investigate the stability of this promising half-metallic FM structure, we compare the 
relative energies of non spin-polarized, ferromagnetic and antiferromagnetically ordered structures 
(Fig. 4).
A G-type AFM solution was obtained in all cases when the moments on the Fe sites were initialized to 
C, A or G-type ordering and non-constrained calculations were performed. Projection of the plane wave 
states into the PAW sphere gave a local magnetic moment of 2.4 $\mu_B$ per iron atom at the equilibrium 
lattice constant. This value is intermediate between high-spin and low-spin configurations for 
a nominally Fe$^{3+}$ ion in a tetrahedral crystal field.
Our FM solutions were obtained by fixing a total magnetic moment of 2.75 $\mu_B$ per Fe atom; this
value is close to that obtained in our local projections for the AFM solutions, and corresponds to
the value obtained when a FM arrangement with initial magnetic moments of 3$\mu_{B}$ per Fe is allowed
to relax to its local minimum. The ``paramagnetic'' results are obtained from a non-spin-polarized GGA calculation.
We find that the antiferromagnetic or non-magnetic solutions are lower in energy than the 
ferromagnetic over the entire volume range studied. The crossover between the AFM and non-magnetic
solutions corresponds to the volume collapse show earlier in Fig. 2(b). 

\begin{figure}
 \centering
 \includegraphics[scale=0.5,bb=0 0 664 528, width=8cm, keepaspectratio=true]{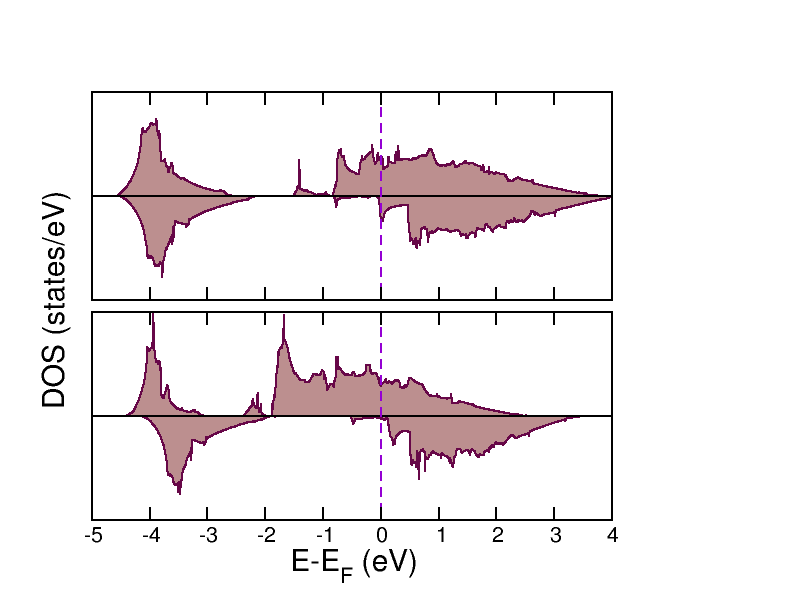}
 % fin_fmdosfeas.png: 0x0 pixel, 0dpi, 0.00x0.00 cm, bb=
 \caption{Spin-polarized density of states for hypothetical FM ordered zincblende FeAs. The Fermi level is set to $0$ eV. Calculated at (a) the equilibrium volume of $38.5$ \AA$^{3}$ per 2-atom formula 
unit, and (b) at the experimental GaAs volume of $45$ \AA$^{3}$.}
 \label{Fig. 3}
\end{figure}

\begin{figure}
 \centering
 \includegraphics[ width=12cm, keepaspectratio=true]{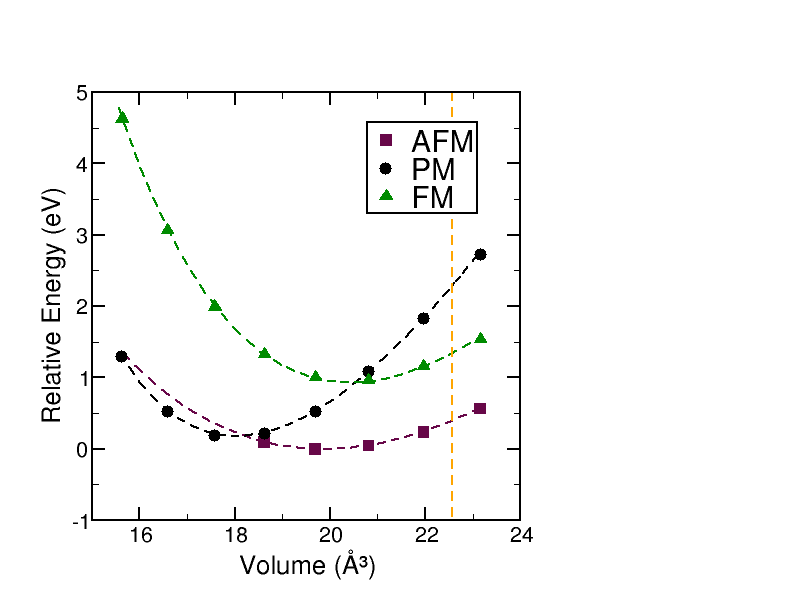}
 % fin_znsall.png: 0x0 pixel, 0dpi, 0.00x0.00 cm, bb=
 \caption{Relative energies of different magnetic orderings in zincblende FeAs as a function of volume of a 2-atom formula unit. The FM curve was calculated by constraining the total moment to 11$\mu_{B}$ for 4 Fe ions. Antiferromagnetic ordering becomes stable above $V=35$ \AA$^{3}$, and remains more stable than FM with cell expansion. The volume corresponding to the GaAs equilibrium is shown by the dashed line.}
 \label{Fig. 4}
\end{figure}

\subsection{Properties of bulk Fe$_{2}$As in Antifluorite and $Cu_{2}Sb$ structural variants}

The Fe-pnictide superconductors consist of a layer of Fe$_{2}$As in the antifluorite structure 
sandwiched between a plethora of other constituents.
Antifluorite (as opposed to fluorite because the anion and cation positions are exchanged) is made up of three interpenetrating face-centred cubic sublattices 
with the Fe(I) sublattice shifted by ($\frac{1}{4}, \frac{1}{4}, \frac{1}{4}$) and the Fe(II) sublattice shifted by ($\frac{1}{4}, \frac{3}{4}, \frac{1}{4}$) along the (111) direction with respect to the As
 lattice at the origin. It can also be considered as `stuffed zincblende' with four interstitial zincblende sites occupied with extra Fe atoms. The Fe and
 As atoms are tetrahedrally coordinated forming edge-shared tetrahedra as shown in Fig 5.

In its ground state, bulk Fe$_{2}$As adopts the \textit{Cu$_{2}$Sb}-type (\textit{C38}) structure 
shown in Fig. 6(a)\cite{Adachi:1988}. There are two Fe cation sites, Fe(I) and Fe(II), with each $a-b$ plane containing a single cation type. These Fe(I) and Fe(II) planes are then alternated in the $c$-direction. Half of the Fe ions are tetrahedrally coordinated with As, and the others form octahedra with six neighbouring As ions. These tetrahedra and octahedra are stacked to form an edge-sharing array.

\begin{figure}
 \centering
 \includegraphics[scale=0.5,bb=0 0 840 839, width=6cm, keepaspectratio=true]{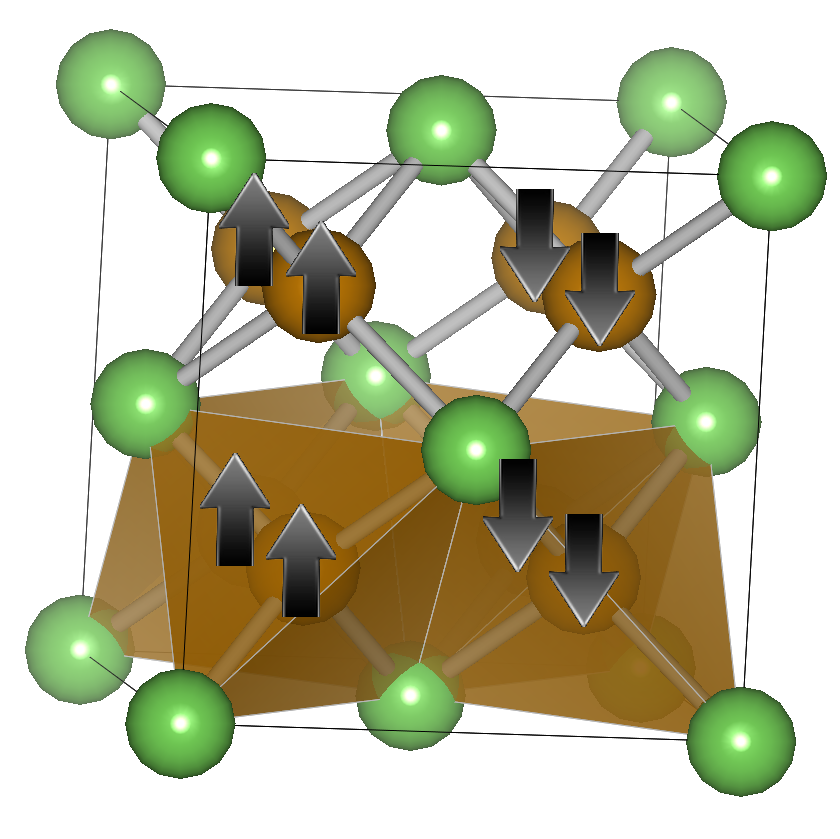}
 % fin_znsall.png: 0x0 pixel, 0dpi, 0.00x0.00 cm, bb=
 \caption{\label{8} Fe$_{2}$As in the antifluorite structure. The calculated ground state magnetic ordering is shown with arrows.}
 \label{Fig. 8}
\end{figure}

Fig. 7 shows the results of our energy-volume calculations for antifluorite and \textit{Cu$_2$Sb}-type 
Fe$_2$As. The \textit{Cu$_2$Sb}-type structure was found to be stable over the whole range of volumes 
calculated, with antifluorite Fe$_2$As less stable by 0.98 eV per formula unit compared to the ground 
state \textit{Cu$_2$Sb}-type structure at its lowest-energy volume. Our calculated structural parameters 
for the \textit{Cu$_2$Sb}-type Fe$_2$As are compared with the experimental values in Table I.

For \textit{Cu$_2$Sb}-type Fe$_2$As, DFT correctly obtains the experimentally determined magnetic ordering 
shown in Fig. 6(b)\cite{Katsuraki:1964}. This is a tri-layer A-type magnetic ordering consisting of three 
layers of FM-coupled Fe atoms coupled antiferromagnetically to the next three FM layers; the measured
T$_{N}$ is 353K. The values of magnetic moment on the two inequivalent Fe sites were calculated to be 
1.25$\mu_{B}$, and 2.18$\mu_{B}$, very close to the experimental values of 1.28$\mu_{B}$ and 2.05$\mu_{B}$ 
\cite{Kaneko:1992}. This good agreement between DFT-GGA and experiment is significant because the 
magnitudes of the magnetic moments in pnictide superconductors are notoriously poorly reproduced by
most flavors of density functional theory. The next lowest energy magnetic ordering that we calculated 
was FM, with a destabilization energy of $1.00$ eV per formula unit.
\begin{figure}
 \centering
 \includegraphics[scale=0.5,bb=0 0 1267 901, width=8cm, keepaspectratio=true]{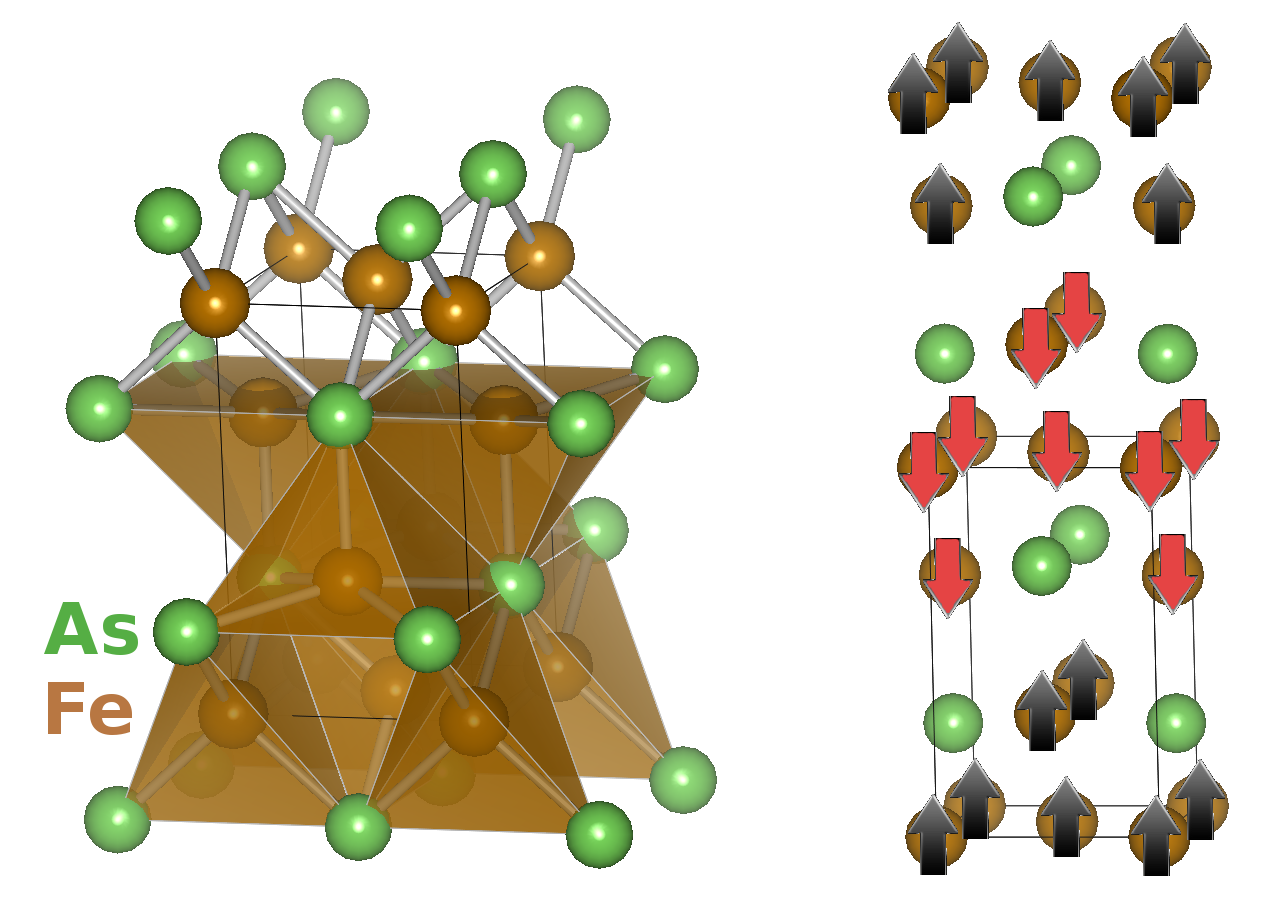}
 % fin_znsall.png: 0x0 pixel, 0dpi, 0.00x0.00 cm, bb=
 \caption{\label{8}(a) Structure of Fe$_2$As in the \textit{Cu$_{2}$Sb} structure. (b) Ground state magnetic ordering of \textit{Cu$_{2}$Sb}-type Fe$_{2}$As.}
 \label{Fig. 8}
\end{figure}

The calculated ground state magnetic ordering for hypothetical bulk antifluorite Fe$_{2}$As was found 
to be A-type with striped antiferromagnetically ordered single $a-b$ plane layers coupled ferromagnetically in 
the $c$-direction as shown in Fig. 5. Again FM was the next most stable ordering, with a destabilization
energy in this case of only 0.17 eV per formula unit at the lowest energy volume.

\begin{figure}
 \centering
 \includegraphics[scale=0.5,bb=0 0 574 510, width=8cm, keepaspectratio=true]{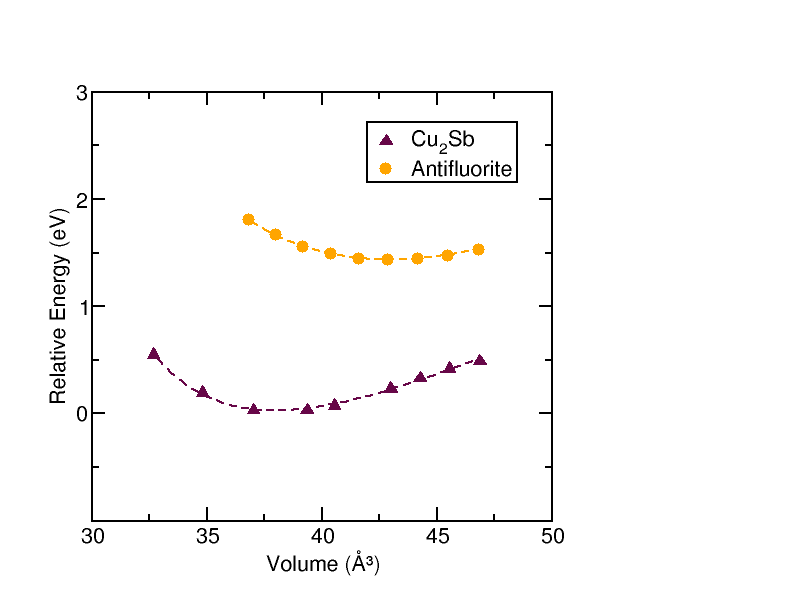}
 % fin_znsall.png: 0x0 pixel, 0dpi, 0.00x0.00 cm, bb=
 \caption{\label{8} Calculated energy versus volume (of one 3-atom formula unit) for the \textit{Cu$_{2}$Sb}-type 
and antifluorite Fe$_{2}$As. The volume was varied by uniform scaling of the calculated equilibrium lattice 
parameters. The \textit{Cu$_{2}$Sb}-type structure is stable across the whole range of volumes studied.}
 \label{Fig. 8}
\end{figure}

\section{Results -- F\lowercase{e}A\lowercase{s}/G\lowercase{a}A\lowercase{s} Superlattices}

\subsection{Zinblende FeAs/ Zincblende GaAs: Spintronic Properties}

While our calculations indicate that ferromagnetic order is not stable for bulk zincblende FeAs, we now 
consider whether it can be stabilized in thin films. We studied superlattices with alternating layers of zincblende structure
FeAs and GaAs (see Fig. 8). We relaxed the structures within the constraint of keeping 
the in-plane lattice parameter fixed at $5.65$ \AA, the experimental lattice 
parameter of GaAs. In all cases we obtained a checkerboard AFM ground state. Our calculated magnetic moments and relative total 
energies for different magnetic orderings for superlattices with 
different numbers of FeAs and GaAs layers are summarized in Table II.

\begin{figure}
 \centering
 \includegraphics[scale=0.5,bb=0 0 2548 1202, width=8cm]{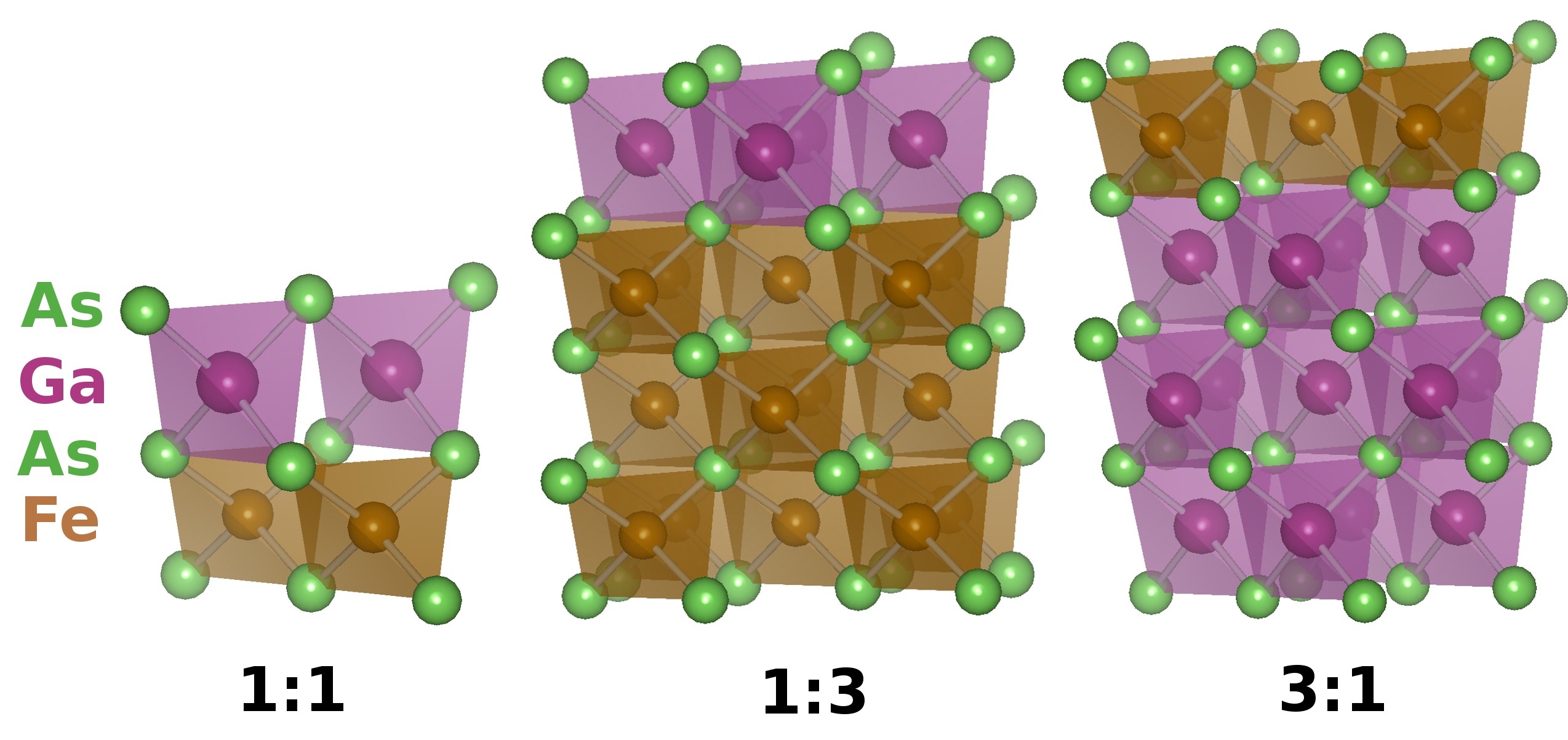}
 % heter1.png: 0x0 pixel, 0dpi, 0.00x0.00 cm, bb=
 \caption{Zincblende structure (FeAs):(GaAs) heterostructures studied in this work.}
 \label{Fig. 6}
\end{figure}

For all of the calculated structures, the ground state magnetic ordering is checkerboard antiferromagnetic. Increasing the ratio of FeAs layers to GaAs does not have an impact on the 
relative stability of the AFM order with repect to the FM order. The values of the magnetic moments are in the range of 2-3$\mu_{B}$ per Fe, which is very similar to that of bulk zincblende FeAs.
The value of the magnetic moment increases as more layers of GaAs are added and seems to correlate with an increase in the out-of-plane Fe-As bond-length as the number of GaAs lazers is increased. To confirm this correlation we repeated the calculation with the Fe, Ga and As ions in the ideal zincblende positions. Comparing the resulting magnetic moment magnitudes of the frozen bond-lengths with those obtained by relaxing the ions shows that the increase in magnitude is not solely a result of quantum confinement, but is caused primarily by the larger Fe-As distance.

\begin{center}
% use packages: array
\begin{table*}
\caption{\label{1} The relative energies of different magnetic orderings (antiferromagnetic, ferromagnetic and paramagnetic) in bulk zincblende FeAs, and in zincblende FeAs/GaAs superlattices fixed to the experimental GaAs lattice parameter. All energies are per Fe atom, and are relative to the ground state AFM order. The values of the Fe magnetic moments in the last column are those for the equilibrium AFM structure}

\begin{tabular}{|l||c|c|c||c|}
\hline
 & E$_{AFM}$ per Fe & E$_{FM}$ per Fe & E$_{PM}$ per Fe & Magnetic Moment per Fe\\ 
 & (eV) & (eV) & (eV) & ($\mu_{B}$) \\ \hline
 FeAs & 0 & 0.174 & 0.209 & 2.152 \\ 
(FeAs)$_{2}$(GaAs)$_{2}$ & 0 & 0.181 & 0.482 & 2.89 \\
 (FeAs)$_{2}$(GaAs)$_{6}$ & 0 & 0.010 & 0.345 & 2.82 \\
 (FeAs)$_{4}$(GaAs)$_{8}$ & 0 & 0.246 & 0.323 & 2.68 \\
 (FeAs)$_{6}$(GaAs)$_{2}$ & 0 & 0.115 & 0.202 & 2.52 \\ 
 (FeAs)$_{8}$(GaAs)$_{4}$ & 0 & 0.207 & 0.280 & 2.63 \\ 
   \hline
\end{tabular}
\end{table*}
\end{center}

The densities of states for the FeAs:GaAs configurations of 2:2, 2:6 and 6:2 are compared with that of bulk zincblende 
FeAs in Fig 9. Despite the change in heterostructuring, the densities of states are very similar in the region
surrounding the Fermi level. All except for the 6:2 superlattice are metallic with Fe states crossing the Fermi level. There is a broad band of Fe-\textit{d} states, with peaks at -3eV and 0.5eV. These Fe states hybridize with the As-\textit{p} states from the FeAs layers.
There is only a small contribution from the As-\textit{p} states in the GaAs layers around the Fermi level. 
However, the main change in increasing the number of GaAs layers is to recover the semiconducting nature of 
the bulk GaAs material. A gap opens up in the (FeAs)$\_{2}$(GaAs)$\_{6}$ heterostructure as shown in the last panel of Fig. 9. 

In summary for this section, all of the zincblende structure FeAs/GaAs heterostructures that we have
studied show robust antiferromagnetic ordering and are therefore unpromising for spintronic applications.
Indeed, their absence of ferromagnetism and/or half-metallicity might explain the experimental difficulties 
associated with spin injection across the Fe-GaAs interface.  

 \begin{figure}
 \centering
\includegraphics[scale=0.5,bb=0 0 654 510, width=9cm,keepaspectratio=true]{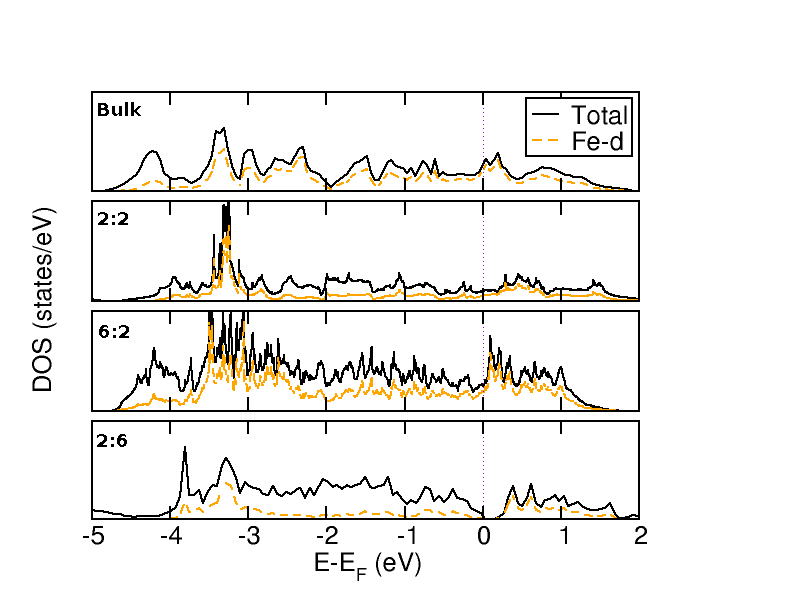}
 \caption{\label{2} Calculated densities of states for zincblende-structure FeAs:GaAs heterostructures. 
The Fermi level is set to 0 eV in each plot and is indicated by the dotted line. The top panel
shows bulk zincblende FeAs. Below this are (FeAs)$_2$(GaAs)$_2$, (FeAs)$_6$(GaAs)$_2$, and the bottom (FeAs)$_2$(GaAs)$_6$. The total DOS is given by the solid black
line. The orbital projected DOS for the Fe-\textit{d} is given by the dashed orange line. }
 \end{figure}

\subsection{Antifluorite Fe$_{2}$As/ Zincblende GaAs: Possible Superconducting Behavior}

The FeAs-based superconducting materials all have a signature nested Fermi surface comprising holes and 
pockets at the Fermi level, resulting from their isolated antifluorite Fe$_2$As layers. Here we investigate
whether such a Fermi surface can be reproduced in antifluorite Fe$_{2}$As / zincblende GaAs heterostructures. 
We studied two specific superlattices, both with one layer of FeAs alternating with a single layer or two layers
of GaAs. Note that, while the bulk antifluorite structure has the formula Fe$_{2}$As, a 1:1 Fe:As ratio is
obtained by taking a slice of the unit cell to reproduce the structure found in the superconducting compounds. 

\begin{figure}
 \centering
 \includegraphics[scale=0.5,bb=0 0 1599 1092, width=8cm]{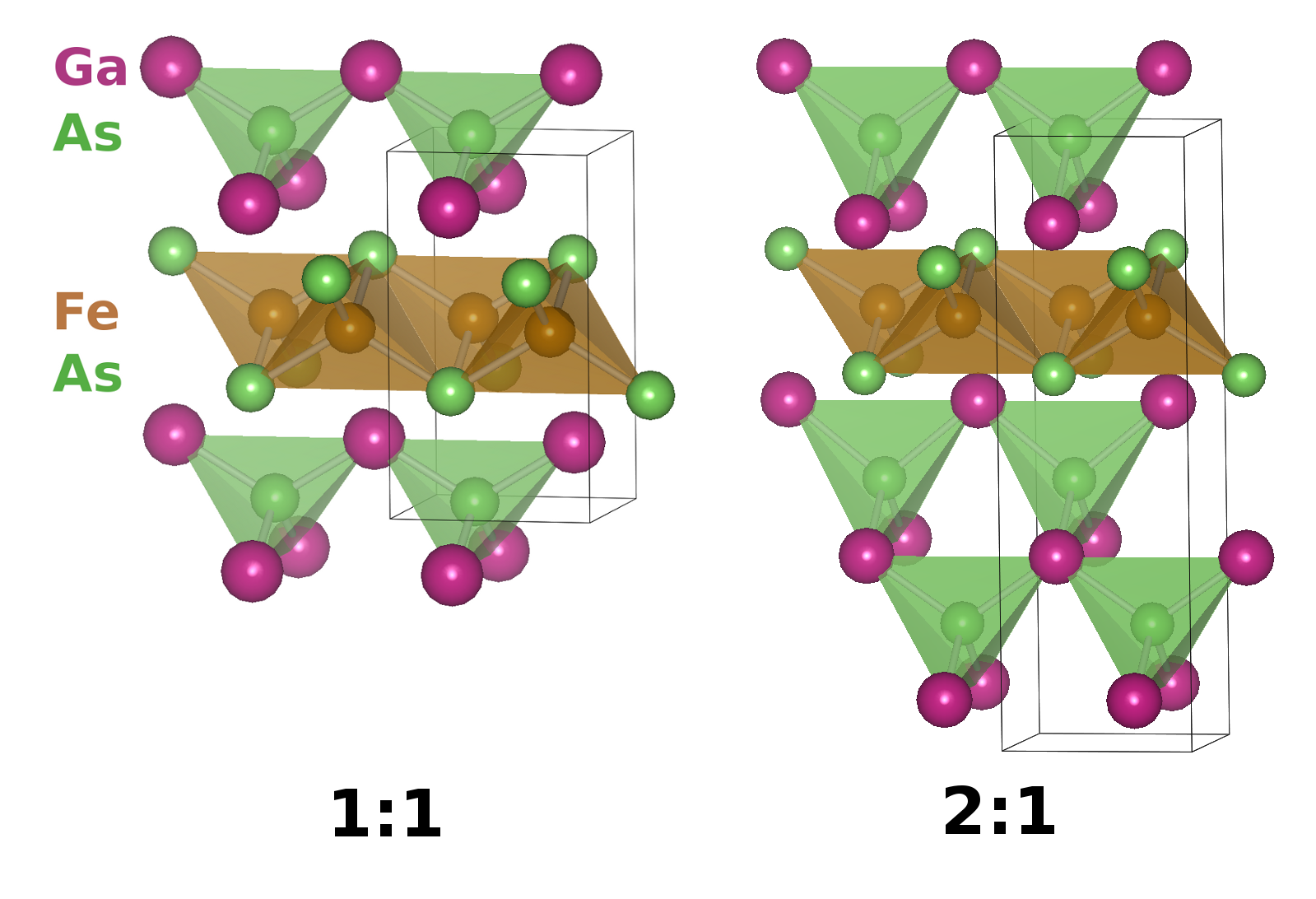}
 % heter1.png: 0x0 pixel, 0dpi, 0.00x0.00 cm, bb=
 \caption{(GaAs):(FeAs) heterostructures for antifluorite FeAs on zincblende GaAs.}
 \label{Fig. 6}
\end{figure}

First we calculate the equilibrium structures of the two heterostructures. As before, 
the in-plane lattice parameter was held fixed at the experimental GaAs lattice parameter 
of $5.65$ \AA, and the out-of-plane lattice parameter and internal coordinates were relaxed 
with this constraint. In Table III we summarize our calculated Fe-As bondlengths and As-Fe-As
bond angles; the latter have been previously shown to correlate with the superconducting transition 
temperature in the parent pnictide compounds, with angles closest to the ideal tetrahedral angle 
of 109.47$^\circ$ yielding the highest $T_c$s\cite{bondangle}. Of the two heterostructures studied, we find that our 1:2 structure has an As-Fe-As bond angle closest to the ideal tetrahedral angle
as a result of the slightly larger FeAs distance in the bilayer structure. 

\begin{center}
% use packages: array
\begin{table}
\caption{\label{1} Calculated FeAs bondlengths and angles for antifluorite FeAs / zinblende GaAs superlattices.} 
\begin{tabular}{|l||c|c|}

\hline
 & Fe-As bond length & As-Fe-As bond angle \\ 
 & (\AA) & (degrees)  \\ \hline
 Fe:Ga 1:1 & 2.36 & 106.56 \\ 
           &       & 115.47 \\ \hline
 Fe:Ga 1:2 & 2.38 & 107.08  \\
           &       & 114.27 \\
   \hline
\end{tabular}
\end{table}
\end{center}

Next we evaluate the magnetic properties of the superlattices. We find that the 
antifluorite FeAs layer maintains several of its bulk magnetic characteristics, as
well as those of the Fe-pnictide superconductors. The ground state magnetic ordering 
for both of the heterostructures was found to be striped antiferromagnetic, which is 
the same as both bulk Fe$_2$As and the Fe-pnictide superconducting parent compounds. 
The average calculated value of the magnetic moment on the Fe sites is 1.6$\mu_{B}$, whereas for the Fe-pnictide compounds the calculated value varies between 1-2 $\mu_{B}$. However, experimentally the Fe magnetic moment value is less than half of these values. This known discrepancy between DFT and experiment is propsed to be a result of spin fluctuations and/or the ambiguity of precisely defining the magnetic form factor in a metal with itinerant magnetism. In view of this, we also expect our calculated values of the magnetic moment to be larger than those experimentally determined.

\begin{figure}
\centering
 \includegraphics[scale=0.5,bb=0 0 951 759, width=8cm,keepaspectratio=true]{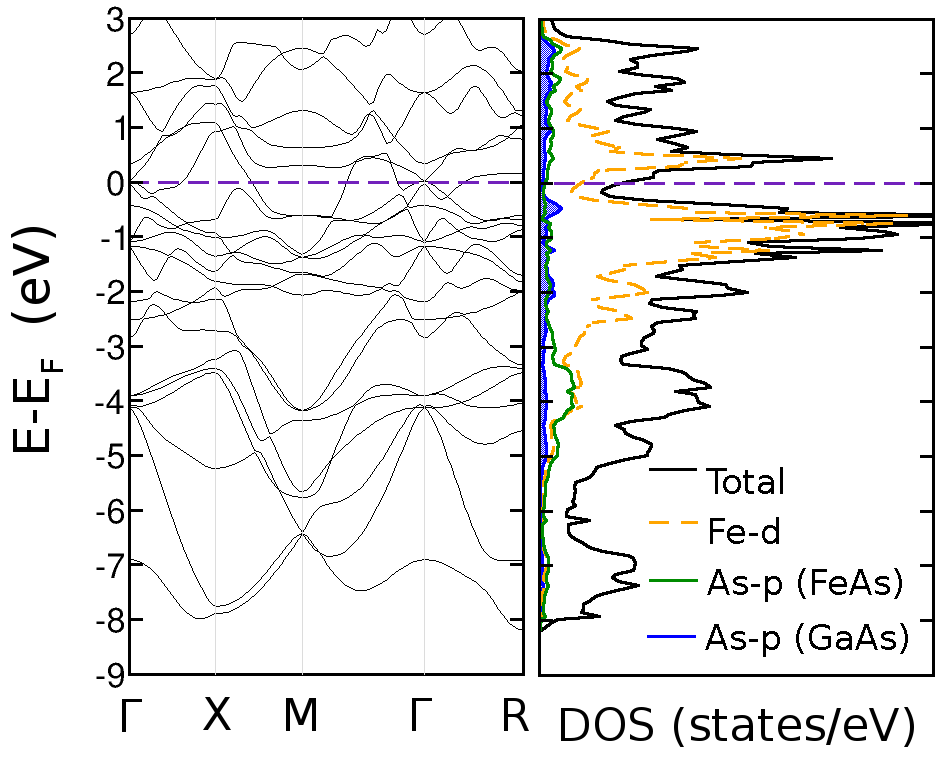}
 \includegraphics[scale=0.5,bb=0 0 951 759, width=8cm,keepaspectratio=true]{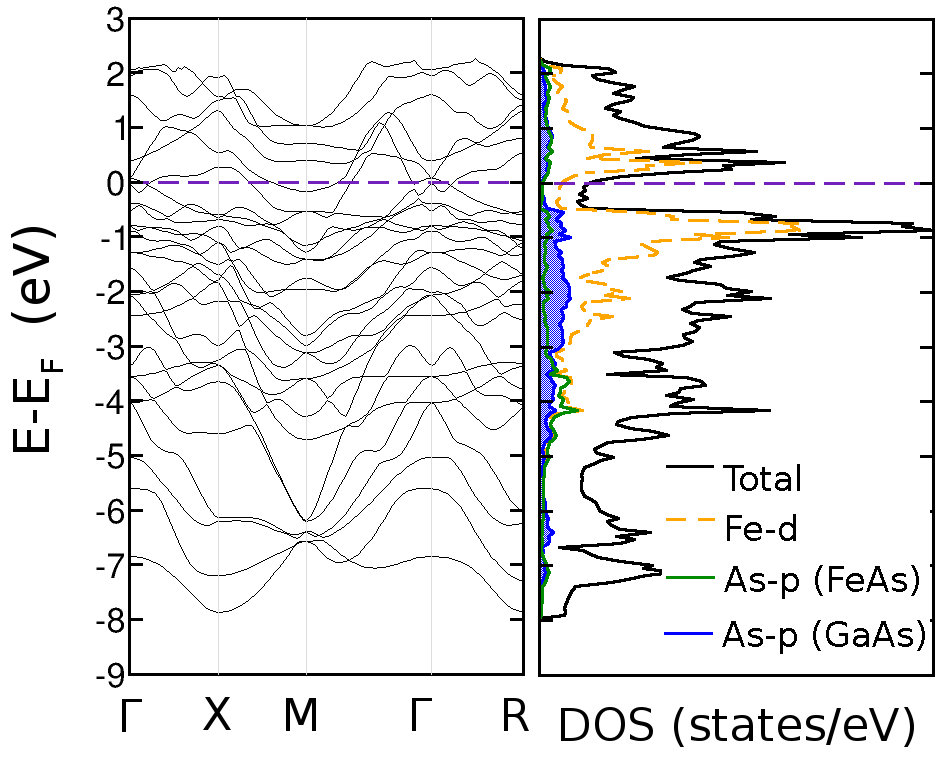}
 \caption{\label{3}Band structures and densities of states for paramagnetic antifluorite Fe$_2$As/ zincblende GaAs superlattices. In all plots, the Fermi level is set to 0 eV. The total densities of states are
given by the black line. The orbitally-projected states of Fe-\textit{d}, As-\textit{p} (from the FeAs layers) and As-\textit{p} (from the GaAs layers) are shown by the dashed orange, green, 
and shaded blue lines respectively. (a) Single layer FeAs with single layer GaAs. (b) Single layer FeAs with double layer GaAs.}
 \end{figure}

Finally, we calculated the electronic structures and Fermi surfaces of the two heterostructures. 
The band structures and densities of states of the two heterostructures in their paramagnetic
state are shown in Fig 11. In both cases we find electron and hole pockets in the 
$\Gamma$ and M directions, characteristic features of the Fe-pnictide superconducting parent compounds. 

As in the pnictide superconductors, the regions surrounding the Fermi level comprise Fe-\textit{d} 
and As-\textit{p} states.  In contrast, however, we find contributions close to the Fermi level from 
components other than the Fe-As layers, in particular from As-\textit{p} states derived from the 
GaAs layers at -0.5 eV. This contribution increases with the addition of more layers of GaAs.

 \begin{figure}
 \centering
\includegraphics[scale=0.5,bb=0 0 1137 2004, width=8cm,keepaspectratio=true]{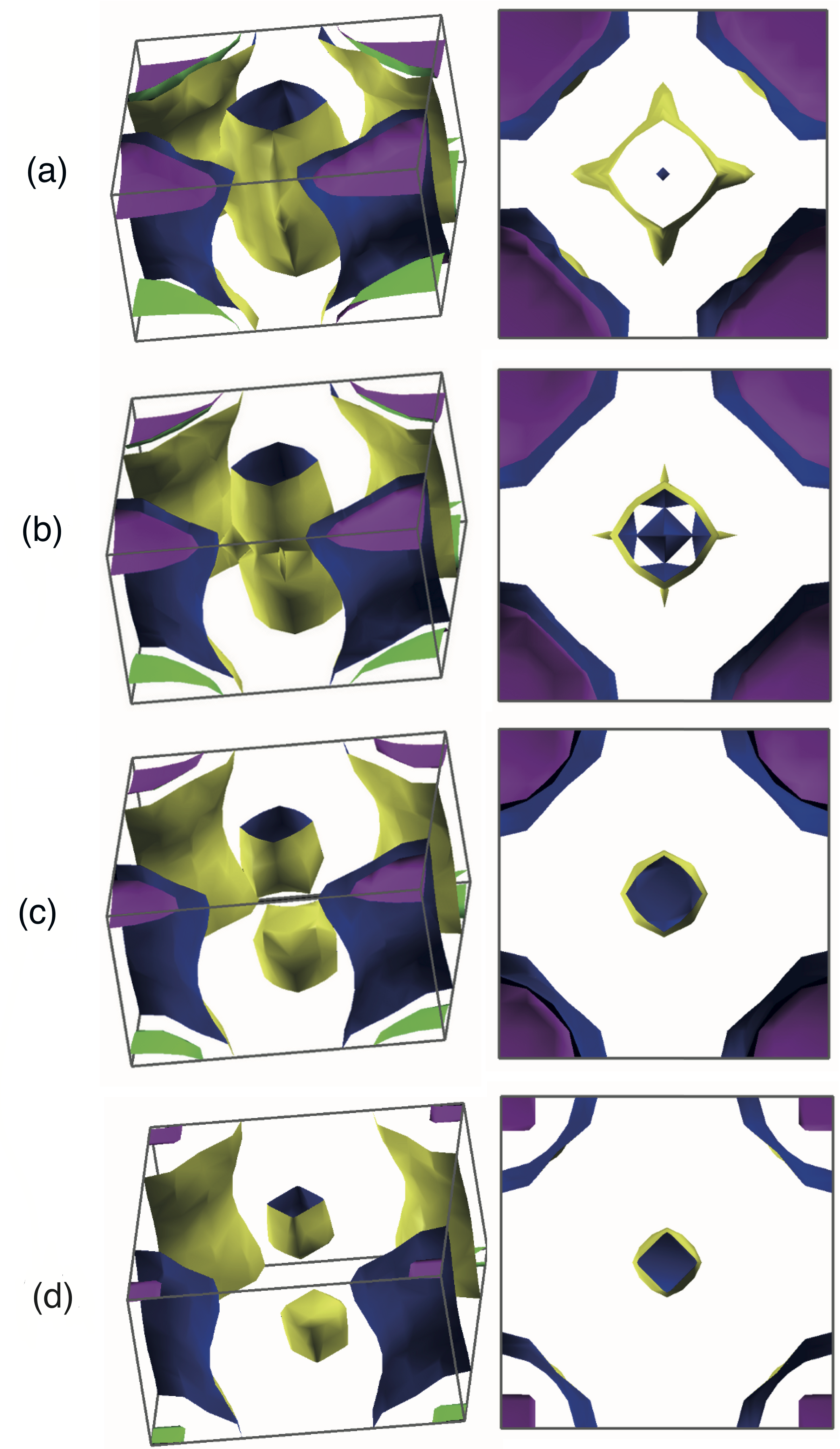}
 \caption{\label{3}Fermi surfaces of FeAs/GaAs heterostructures for various levels of electron doping relative to the Fermi level. 
(a) E$_F$ (b) E$_F$ - 0.1eV (c) E$_F$ - 0.2eV (d) E$_F$ - 0.3eV}
 \end{figure}

In Fig. 12 we show our calculated Fermi surface, and its varation with doping using a rigid band model, 
for the heterostructure with double layers of GaAs. 
The Fermi surface of the FeAs/GaAs heterostructure shares some features with the 
superconducting Fe-pnictide surfaces: At E$_F$, the centre of the Brillouin zone consists
of a distorted cylinder with a sphere nested inside. The zone edges have another cylinder 
with pockets at the corners. Reducing the electron count, the inner sphere disappears and 
the cylinders begin to shrink. At an energy of 0.2 eV below the Fermi level, the central 
cylinder splits into two cylindrical shapes. Unlike the superconducting pnictides, however, 
single cylinders rather than double cylinders occur at the $\Gamma$ and M points.
Therefore, the characteristic Fe-pnictide nesting between the zone centre and zone edges cannot occur.
In addition the surface is manifestly three dimensional, whereas that of the superconductors
is two dimensional.

\section{Summary}

We studied two types of FeAs/GaAs heterostructures, both with GaAs in the zincblende structure,
and with FeAs in the antifluorite structure or the zincblende structure. 

The zincblende/zincblende heterostructures allowed us to investigate possible spintronic applications
of FeAs/GaAs. Our calculations suggest that desirable half-metallicity and/or ferromagnetism are unlikely,
however. Instead, the robustly stable AFM ordering offers a possible explanation for the failure of 
spin injection across the Fe/GaAs interface.

Antifluorite Fe$_{2}$As/GaAs heterostructures would allow direct integration of superconductors
into III-V semiconductor technologies if they reproduced the key features of the Fe-pnictide 
superconductors. While we see several features in common with the ferropnictide superconductors
-- the same AFM ordering, Fe $d$ and As $p$ states around the Fermi energy and some similarities in
the positions of the hole pockets in Fermi surface -- we find that covalent bonding with the 
As-\textit{p} states from the GaAs layers causes the Fermi surface to be unfavorably three-dimensional
and inhibits the electron-hole pocket nesting that is characeterestic of the Fe-pnictide superconductors. 
A possible route to avoid this Fermi level mixing is to use a more dissimilar spacer layer than GaAs such 
as GaN.

\section{Acknowledgements}
We would like to thank Kris Delaney and James Rondinelli for helpful discussions. 
This work was supported by the MRSEC Program of the National Science Foundation under Award No. DMR05-20415 and ETH Z\"{u}rich. We also made use of the CNSI Computing Facility (Hewlett Packard) and Teragrid.

\bibliography{july}

\end{document}